# Exact Ionization Potentials from Wavefunction Asymptotics: The Extended Koopmans' Theorem, Revisited

Diederik Vanfleteren[1], Dimitri Van Neck[1], Paul W. Ayers[2], Robert C. Morrison[3], and Patrick Bultinck[4]

1. Center for Molecular Modeling, Ghent University, Proeftuinstraat 86, B-9000 Gent, Belgium
2. Department of Chemistry; McMaster University; Hamilton ON Canada L8S 4M1
3. Department of Chemistry, East Carolina University, Greenville NC 27858, USA
4. Department of Inorganic and Physical Chemistry, Ghent University, Krijgslaan 281(S3), 9000 Gent, Belgium

A simple explanation is given for the exactness of the extended Koopmans' theorem for computing the removal energy of any many-electron system to the lowest-energy ground state ion of a given symmetry. In particular, by removing the electron from a "removal orbital" of appropriate symmetry that is concentrated in the asymptotic region, one obtains the exact ionization potential and the exact Dyson orbital for the corresponding state of the ion. It is argued that the EKT is not restricted to many-electron systems but holds for any finite many-body system, provided the interaction vanishes for increasing interparticle distance. A necessary and sufficient condition for the validity of the extended Koopmans' theorem for any state (not just the lowest-energy states of a given symmetry) in terms of the third-order reduced density matrix is stated and derived.

# I. The Extended Koopmans' Theorem

As an electron in an *N*-electron system moves further and further away, the wavefunction of the system begins to resemble an *N*-1 electron system, polarized by the distant electron. In the asymptotic limit,

$$\lim_{|\mathbf{r}_N|\to\infty} \Psi^{(N)}\left(\mathbf{r}_1,s_1;\mathbf{r}_2,s_2;\ldots;\mathbf{r}_N,s_N\right) \propto \Psi_s^{(N-1)}\left(\mathbf{r}_1,s_1;\mathbf{r}_2,s_2;\ldots;\mathbf{r}_{N-1},s_{N-1}\right) \tag{1}$$

where $\Psi_s^{(N-1)}$ is the lowest-energy state of the *N-1* electron system that is consistent with the spin- and spatial- symmetry states of the original *N*-electron state ($\Psi^{(N)}$) and the electron that is removed. Thus any eigenfunction of an *N*-electron system contains within it *exact* information about the low-energy eigenfunctions of the *N-1* electron system. The extended Koopmans' theorem (EKT) is a practical numerical procedure for extracting this exact information.[1-3] The EKT often provides reasonably accurate approximations for some of the higher-energy states of the *N-1* electron system also.[4-11]

The extended Koopmans' theorem is derived from the orbital representation of the electron-removal process in Eq. (1),[1-3]

$$\Psi_k^{(N-1)}\left(\mathbf{z}_1;\ldots;\mathbf{z}_{N-1}\right) = \int \phi_k\left(\mathbf{z}_N\right)\Psi^{(N)}\left(\mathbf{z}_1;\ldots;\mathbf{z}_N\right)d\mathbf{z}_N , \tag{2}$$

where $\mathbf{z}=(\mathbf{r},s)$ denotes both the spatial- and spin-coordinates of the electrons. The EKT removal orbital, $\phi_k$, is determined by minimizing the energy of the *N-1* electron state or, equivalently, optimizing the ionization energy

$$\begin{aligned}\mathcal{I}_k &= E_k^{(N-1)} - E_0^{(N)} \\ &= \frac{\left\langle \int\phi_k(\mathbf{z}_N)\Psi^{(N)}(\mathbf{z}_1\ldots\mathbf{z}_N)d\mathbf{z}_N \middle| \int\phi_k(\mathbf{z}_N)\left(\hat{H}^{(N-1)}-\hat{H}^{(N)}\right)\Psi^{(N)}(\mathbf{z}_1\ldots\mathbf{z}_N)d\mathbf{z}_N \right\rangle_{1\ldots N-1}}{\left\langle \int\phi_k(\mathbf{z}_N)\Psi^{(N)}(\mathbf{z}_1\ldots\mathbf{z}_N)d\mathbf{z}_N \middle| \int\phi_k(\mathbf{z}_N)\Psi^{(N)}(\mathbf{z}_1\ldots\mathbf{z}_N)d\mathbf{z}_N \right\rangle_{1\ldots N-1}}\end{aligned} \tag{3}$$

This gives the EKT equation,

$$\int \phi_k(\mathbf{z}_N)\left(-\frac{\nabla_N^2}{2} + v_{ext}(\mathbf{z}_N) + \sum_{i=1}^{N-1}\frac{1}{|\mathbf{r}_i - \mathbf{r}_N|}\right)\Psi^{(N)}(\mathbf{z}_1,\ldots\mathbf{z}_N)d\mathbf{z}_N = -\varepsilon_k \int \phi_k(\mathbf{z}_N)\Psi^{(N)}(\mathbf{z}_1,\ldots\mathbf{z}_N)d\mathbf{z}_N . \quad (4)$$

The orthonormality of the *N*-1 electron wavefunctions defined in Eq. (2) implies that the EKT removal orbitals to be orthonormal in the metric of the one-electron reduced density matrix,

$$\iint \phi_k(\mathbf{z})\gamma(\mathbf{z},\mathbf{z}')\phi_l(\mathbf{z}')d\mathbf{z}d\mathbf{z}' = \delta_{kl} . \quad (5)$$

The EKT equation is more commonly written in terms of density matrices, either in the spatial representation,[3,12]

$$\begin{aligned}
\iint \phi_k(\mathbf{z}_1')&\left(-\frac{\nabla_1^2}{2} + v_{ext}(\mathbf{z}_1) + \frac{(N-1)}{|\mathbf{r}_1 - \mathbf{r}_2|}\right)\Gamma_2^{(N)}(\mathbf{z}_1,\mathbf{z}_2;\mathbf{z}_1',\mathbf{z}_2)d\mathbf{z}_2 d\mathbf{z}_1' \\
&= -\varepsilon_k \iint \phi_k(\mathbf{z}_1')\Gamma_2^{(N)}(\mathbf{z}_1,\mathbf{z}_2;\mathbf{z}_1',\mathbf{z}_2)d\mathbf{z}_2 d\mathbf{z}_1' \\
\int \phi_k(\mathbf{z}_1')&\left(-\frac{\nabla_1^2}{2} + v_{ext}(\mathbf{z}_1)\right)\Gamma_1^{(N)}(\mathbf{z}_1;\mathbf{z}_1')d\mathbf{z}_1' + 2\iint \phi_k(\mathbf{z}_1')\left(\frac{1}{|\mathbf{r}_1 - \mathbf{r}_2|}\right)\Gamma_2^{(N)}(\mathbf{z}_1,\mathbf{z}_2;\mathbf{z}_1',\mathbf{z}_2)d\mathbf{z}_2 d\mathbf{z}_1' \\
&= -\varepsilon_k \int \phi_k(\mathbf{z}_1')\Gamma_1^{(N)}(\mathbf{z}_1;\mathbf{z}_1')d\mathbf{z}_1'
\end{aligned} \quad (6)$$

or the orbital representation,[1,2,9]

$$\sum_\lambda \left\langle \Psi^{(N)}\left|a_\lambda^+[a_\kappa,\hat{H}]\right|\Psi^{(N)}\right\rangle c_{k\lambda} = -\varepsilon_k \sum_\lambda \left\langle \Psi^{(N)}\left|a_\lambda^+ a_\kappa\right|\Psi^{(N)}\right\rangle c_{k\lambda} \quad (7)$$

$$\sum_\lambda \left(\sum_\mu h_{\kappa\mu}\gamma_{\lambda\mu} + \sum_{\mu\nu o} V_{\kappa\mu\nu o}\Gamma_{\lambda\mu\nu o}\right)c_{k\lambda} = \varepsilon_k \sum_\lambda \gamma_{\kappa\lambda} c_{k\lambda} \quad (8)$$

where $|\phi_k\rangle = \sum_\lambda c_{k\lambda} a_\lambda^+ |0\rangle$ is the expression for the EKT removal orbital and the reduced density matrices are defined as

$$\gamma_{\lambda\mu} = \left\langle \Psi^{(N)}\left|a_\mu^+ a_\lambda\right|\Psi^{(N)}\right\rangle \quad (9)$$

$$\Gamma_{\lambda\mu\nu o} = \left\langle \Psi^{(N)} \middle| a_\nu^+ a_o^+ a_\mu a_\lambda \middle| \Psi^{(N)} \right\rangle \tag{10}$$

Finally, the EKT can be expressed in terms of quantities related to electron propagator theory as,

$$\int M_1^{(-)}(\mathbf{z},\mathbf{z}')\phi_k(\mathbf{z}')d\mathbf{z}' = \varepsilon_k \int M_0^{(-)}(\mathbf{z},\mathbf{z}')\phi_k(\mathbf{z}')d\mathbf{z}' \tag{11}$$

where

$$\begin{aligned} M_n^{(-)}(\mathbf{z},\mathbf{z}') &= \frac{1}{2\pi i}\int E^n G(\mathbf{z},\mathbf{z}',E)e^{i\eta E}dE \\ &= \sum_k (-\mathcal{I}_k)^n g_k(\mathbf{z})g_k(\mathbf{z}') \end{aligned} \tag{12}$$

are the energy moments of the electron propagator that are associated with electron removal. The $k^{\text{th}}$ Dyson orbital for electron removal is denoted $g_k = \left\langle \Psi_k^{(N-1)} \middle| \Psi^{(N)} \right\rangle_{1\ldots N-1}$ and the associated exact removal energy is denoted $\mathcal{I}_k = E_k^{(N-1)} - E_0^{(N)}$. The zeroth energy moment is just the one-electron reduced density matrix, $M_0^{(-)} = \gamma$.

In all of these expressions, the quantity

$$-\varepsilon_k \approx \mathcal{I}_k \tag{13}$$

gives an approximation to the electron removal energy. If one uses the Hartree-Fock wavefunction/density matrices/electron propagator in these quantities, then $\varepsilon_k$ is identical to the Hartree-Fock orbital energy and Eq. (13) is just the usual statement of Koopmans' theorem. So Eqs. (4)-(11) represent an extension of Koopmans' theorem to correlated electronic structure theory calculations.

The EKT is an important tool in electronic structure theory because it provides a simple and computationally efficient way to extract information about the successive ionization potentials of a system from correlated electronic structure methods. Together, the (generally approximate) electron removal energies, $\varepsilon_k$, and their associated approximate Dyson orbitals,

$$g_k^{(EKT)}(\mathbf{z}) = \int \gamma(\mathbf{z},\mathbf{z}')\phi_k(\mathbf{z}')d\mathbf{z}' \quad (14)$$

provide "correlated" analogues that have similar conceptual utility to the single-particle orbitals from independent particle models like Hartree-Fock and Kohn-Sham DFT.[13,14] The analogous orbitals and energies for electron attachment can be defined (just interchange the creation and annihilation operators in Eq. (7))[2] but, like the corresponding quantities in Hartree-Fock theory, they are usually less accurate than the corresponding ionization-related quantities.

The EKT has been used to extract ionization potentials from many different types of correlated calculations. The most straightforward implementation uses the result of CASSCF/MCSCF/full-CI calculations (using GAMESS-US);[9] it is also possible to use the EKT to obtain ionization potentials from Møller-Plesset perturbation theory and QCISD calculations (using Gaussian IOp(6/51))[15]. The EKT has been used to test approximate self-energies in electron propagator approaches.[16] Ionization potentials in density-matrix functional theory can be computed with EKT because the two-electron matrix elements needed to evaluate Eq. (8) are approximated in terms of the one-particle reduced density matrix in such approaches.[17-20] It is clear that the EKT could provide similar information for computational approaches using higher-order reduced density matrices also.[21-29] EKT has also been explored as a method for computing the change in electron density from electron removal/attachment (the so-called Fukui function[30-32]) and other reactivity indicators in DFT-based chemical reactivity theory.[33]

## II. The Exactness of the Extended Koopmans' Theorem

The extended Koopmans' theorem is *exact* for the lowest-energy electron removal energy, $\varepsilon_0$. This remarkable statement has been the subject of much doubt and debate.[7,8,11,12,34-38] For example, it seems remarkable that the lowest ionization energy of any system could be determined *exactly* from using Eq. (11) since $M_1^{(-)}$ is the operator representing the *average* electron removal energy. Indeed, if one compares the EKT result to results from perturbation theory,[34,35] one observes that second-order terms are missing; it is just that these terms seem to be zero for the lowest-energy electron removal process.[36]

The usual argument for the exactness of the EKT uses the fact that the first-order density matrix can be expressed using approximate Dyson orbitals from the EKT,

$$\gamma(\mathbf{z},\mathbf{z}') = \sum_k \left(g_k^{(EKT)}(\mathbf{z}')\right)^* g_k^{(EKT)}(\mathbf{z}). \tag{15}$$

Moreover, by analogy to the argument for the asymptotic decay of the Hartree-Fock orbitals,[39] Morell, Parr, and Levy were able to show that the $r \to \infty$ asymptotic decay of the EKT Dyson orbitals is governed by the smallest EKT ionization potential,

$$g_k^{(EKT)}(\mathbf{z}) \sim e^{-r\sqrt{-2\varepsilon_0}}. \tag{16}$$

Since the $r \to \infty$ asymptotic decay of the density matrix is[40-43]

$$\gamma(\mathbf{z},\mathbf{z}') \sim e^{-r\sqrt{2\mathcal{I}_0}}, \tag{17}$$

it follows that $\varepsilon_o = -\mathcal{I}_0$. The extended Koopmans' theorem is thus exact for the first ionization potential. It should be noted that the derivation of Eq. (17) uses the Coulomb form of interaction between the particles. Whether the EKT works for non-Coulombic interactions is an open problem, but we will argue later that the EKT is very general and give a non-Coulombic example where it is exact.

While the asymptotic argument for the exactness of the EKT is convincing by the standards of mathematical rigor that are typical in theoretical chemical physics, it may be criticized. In particular, it is certainly possible to expand one function (say, $e^{-2r}$) in terms of an infinite number of functions which have a different asymptotic decay (e.g., the orthonormal set, $(4\pi)^{-1/2} L_n^2(r) e^{-r/2}$ where $L_n^2(r)$ are associated Laguerre polynomials). The fact that the asymptotic decay of the density matrix is given by Eq. (17) thus strongly suggests—but does not rigorously prove—that the functions used in the expansion, $g_k^{(EKT)}$, also decay in like manner.

Our assertion is that the EKT holds precisely because of Eq. (1). That is, if the EKT removal orbital, $\phi_k(\mathbf{z})$, is concentrated sufficiently far away from the molecule, then the *N-1* electron wavefunction that results from Eq. (2) will be accurate. Moreover, the further $\phi_k$ moves from the molecule, the more accurate $\Psi_k^{(N-1)}$ becomes. This also implies that the optimal EKT removal orbital should be interpreted as a generalized function (a limit of a sequence of functions): there is no well-defined physical orbital that solves Eq. (4) in the basis set limit.

To confirm this hypothesis numerically, we performed a series of full-configuration interaction calculations on the Beryllium atom using the six *s*-type functions from the cc-pV5z basis and an additional spherically-symmetric radial Gaussian function, $\chi(\mathbf{r}) \sim e^{-5(r-r_0)^2}$. This extra function is a spherical shell that peaks $r_0$ atomic units from the Beryllium nucleus. The value of $r_0$ is successively taken to be 1.0, 7.0, 9.0, or 11.0 atomic units, yielding radial wavefunctions concentrated at increasing distances from the nucleus. Note that the Coulomb integrals involving the added spherical shell orbital are nonstandard; they were calculated using numerical quadrature. The EKT removal orbital that corresponds to the lowest ionization potential is expected to be concentrated as far from the atom as possible. As figure 1 indeed shows, for the more distant spherical shells the EKT removal orbital is nearly proportional to the added radial shell function. We infer that as the basis set approaches completeness, the EKT removal orbital becomes concentrated infinitely far away from the atom or molecule. The cc-pV5Z basis is already sufficiently diffuse to give good results for the ionization potential and the computed lowest IP is almost exact ( $IP_{EKT, r_0 = 1}$ = .29900 a.u.; $IP_{exact}$ = .29898 a.u.) even for the nearest spherical shell ($r_0$ = 1 a.u.). For spherical shells concentrated in the nether regions of the atom, $r_0 \geq 7$ a.u., the lowest EKT IP agrees with the full-CI value to the limits of our numerical precision.

Further confirmation of our interpretation was obtained by revisiting the accurate calculations on the Beryllium atom from ref. [10]. We started with the 6-311G basis for Be (located at the origin) and then extended this basis set by adding twelve ghost atoms at positions $\mathbf{r}_0$, distributed symmetrically along the z-axis at $z_0$ = ±1.0, ±2.0, ... ±6.0 Angstroms. A single primitive Gaussian wavefunction $e^{-|\mathbf{r}-\mathbf{r}_0|^2}$ is centered on each ghost atom. The EKT orbital was then computed from an MCSCF calculation using the first 18 of the 25 orbitals computed from this basis. The generalized eigenvalue problem in Eq. (8) is ill-conditioned when the first-order density matrix is nearly singular, so we eliminated a natural orbital with occupation less than $10^{-11}$ from our the calculations of the EKT orbital. The resulting EKT orbital is plotted in Figure 2; notice that the orbital density is concentrated on the distant ghost atoms. The orbital density is located mainly on the ghost atoms at 4.0 and 5.0 Angstroms instead of the most distant ghost atom (at 6.0 Angstroms) because a natural orbital that is mostly on the furthest ghost atoms is eliminated from the calculation because of its vanishingly small occupation number.

The asymptotic arguments employed by Morrell, Parr, and Levy rely on the Coulombic $r_{ij}^{-1}$ interactions between electrons.[3] Based on the observed differences between the expressions for electron removal energies from the EKT and from order-by-order expansion, it was speculated that perhaps the EKT is only exact for the lowest electron-removal energy in Coulombic systems. However, the argument formulated in the previous paragraphs relies only upon the validity of Eq. (1), which is true for any system, including systems with short-range interparticle potential. The only proviso is that the interaction vanishes with increasing interparticle distance. To ensure that the EKT is indeed true for any system, we considered a self-bound system of bosons interacting via delta-function attractive potentials in one dimension with the Hamiltonian

$$\hat{H} = \frac{-\hbar^2}{2m} \sum_{i=1}^{N} \frac{\partial^2}{\partial x_i^2} - g \sum_{i=1}^{N-1} \sum_{j=i+1}^{N} \delta(x_i - x_j). \tag{18}$$

This model is exactly solvable for arbitrary particle number.[44,45] For all $N \geq 2$ this system has only one bound state, with energy $E_0^{(N)} = \frac{-1}{12} \lambda g N(N^2 - 1)$ where $\lambda = \frac{mg}{2\hbar^2}$.

As in the previous section, we focused our study on the four-particle system, $N = 4$, with ionization potential $\mathcal{I}_0 = 3\lambda g$. In this case, the one-particle reduced density matrix ($\gamma = M_0^{(-)}$) and the removal energy matrix ($M_1^{(-)}$) have the expressions

$$M_0^{(-)}(x, x') = \frac{3\lambda}{5} \begin{pmatrix} 30e^{-3s} - 15e^{-6s+3v} - 9e^{-6s-5v} + 3e^{-9s+6v} + 2e^{-9s-4v} + e^{-9s-10v} \\ +\Theta(v-s)\left(-3e^{-9s+6v} + 3e^{6s-9v} + 15e^{-6s+3v} - 15e^{3s-6v} - 30e^{-3s} + 30e^{-3v}\right) \\ +\Theta(v-3x)\left(10e^{-7v} - 15e^{-3s-6v} + 9e^{-6s-5v} - 3e^{6s-9v} - 2e^{-9s-4v} + e^{9s-10v}\right) \end{pmatrix} \tag{19}$$

$$M_1^{(-)}(x,x') = \frac{\lambda^2 g}{5} \begin{pmatrix} 270e^{-3s} - 135e^{-6s+3v} - 9e^{-6s-5v} + 27e^{-9s+6v} + 18e^{-9s-4v} + e^{-9s-10v} \\ +\Theta(v-s)\left(-27e^{-9s+6v} + 27e^{6s-9v} + 135e^{-6s+3v} - 135e^{3s-6v} - 270e^{-3s} + 270e^{-3v}\right) \\ +\Theta(v-3x)\left(130e^{-7v} - 135e^{-3s-6v} + 9e^{-6s-5v} - 27e^{6s-9v} - 22e^{-9s-4v} + e^{9s-10v}\right) \end{pmatrix} \quad (20)$$

where $v = \lambda|x - x'|$, $s = \lambda|x + x'|$, and $\Theta(x)$ is the Heaviside step function. The exact Dyson orbital is

$$g_0(x) = \sqrt{2\lambda}\left(3e^{-3\lambda|x|} - \tfrac{3}{2}e^{-9\lambda|x|} + \tfrac{3}{10}e^{-15\lambda|x|}\right) \quad (21)$$

In the $x \to +\infty$ asymptotic limits, these quantities have the asymptotic forms

$$M_0^{(-)}(x,x') \sim g_0(x')3\sqrt{2\lambda}e^{-3\lambda x} \quad (22)$$

$$M_1^{(-)}(x,x') \sim \mathcal{I}_0\left(g_0(x')3\sqrt{2\lambda}e^{-3\lambda x}\right) \quad (23)$$

$$g_0(x) \sim 3\sqrt{2\lambda}e^{-3\lambda x} \quad (24)$$

For this self-bound system, characteristic exponential rate of asymptotic decay, $\alpha = 3\lambda$, is related to the ionization potential as $\alpha = \sqrt{2\mu\mathcal{I}_0}$, where $\mu = \frac{mN}{N-1}$ is the reduced mass for the motion of one particle (far away from the others) relative to the center of mass of the other $N-1$ particles.

We substituted Eqs. (19) and (20) into Eq. (11) and solved for the EKT removal orbital and its eigenvalue on a grid of evenly spaced points. The equation to be solved is very ill-conditioned, and it is necessary to eliminate all of the natural orbitals with small eigenvalues. Once this is done, the equation can be solved, and we observe that the EKT removal orbital, $\phi_0(x)$ is concentrated entirely at the edge of the $x$-interval covered by the grid. In accord with our expectations, the EKT removal energy, $\varepsilon_0$, is correct, and the EKT Dyson orbital, $g_0^{(EKT)}(x)$, matches the analytic result from Eq. (24).

It has been repeatedly noted that, for atoms, the EKT seems to be exact not only for the lowest ionization potential, but also for the lowest ionization potential to other symmetry states.[5,7,10,11] We can now explain this finding. By removing an electron with a specified spin and angular momentum (i.e., a specific $m_s$ and $m_\ell$), one can often access excited states of the ion. For example, consider the $N$-electron wavefunction with $M_S^{(N)} = S^{(N)}$. Removing a majority-spin electron will produce a state of the ionized system with $M_S^{(N-1)} = S^{(N)} - 1$; removing a minority-spin electron will produce a state of the ion with $M_S^{(N-1)} = S^{(N)} + 1$. If the ground state of the ionized system has smaller spin-multiplicity than the ground state of the original molecule, then one should obtain *two* ionization potentials exactly: one for electron removal from the spin-majority channel and one for electron removal from the spin-minority channel. In cases where the spin-multiplicity of the ionized system is very different from that of the original molecule, the EKT may fail to give the lowest ionization potential, and instead give exact results only for higher-energy excited states of the cation. These results are similar to the situation in spin-density-functional theory, where the eigenvalues of the highest-occupied orbitals of both α-spin and β-spin equal minus one times the associated electron removal energies.[46-49]

The case of angular momentum symmetry is similar. By choosing the EKT removal orbital with different values of $m_\ell$, one can obtain exact ionization potentials to excited states of the ion that have higher-angular momentum than the ground state.

## III. Necessary And Sufficient Conditions for Exactness

When is the extended Koopmans' theorem exact not only for the removal energy to the lowest state of a given symmetry, but also to certain other, higher-order, excited states? There does not seem to be any easy criterion for this, though it is known that the EKT is accurate (though perhaps not exact) even for ionization to highly excited states of the cation (e.g., the $1s^22s^2 \rightarrow 1s^12s^2$ ionization of Be[10]). Pernal and Cioslowski gave a criterion for all of the EKT removal energies to be exact in terms eigenvalues, $\{v_k\}$, and eigenvectors, $\{\Phi_k\}$, of the *N-1* particle reduced density matrix,

$$\Gamma_{N-1}^{(N)} = \sum_k v_k \Phi_k(\mathbf{z}_1'...\mathbf{z}_{N-1}')\Phi_k(\mathbf{z}_1...\mathbf{z}_N). \tag{25}$$

Specifically, all of the EKT removal energies are exact if for every $j$ such that $v_j > 0$ and $k$ such that $v_k = 0$, the energy matrix element between $\Phi_j$ and $\Phi_k$ vanishes: $\langle \Phi_j | \hat{H}^{(N-1)} | \Phi_k \rangle = 0$.[12] This criterion is unlikely to be satisfied, and is difficult to evaluate because it requires the *N-1* electron reduced density matrix. It is worth recalling that whenever $v_j > 0$, it is equal to one of the natural orbital occupation numbers.[50]

It seems more likely that *some* of the EKT removal energies are exact while others are approximate. In that case, the Pernal-Cioslowski condition will be violated. A simple test that is *necessary and sufficient* for the exactness of any specific EKT removal energy and the associated Dyson orbital is obtained by noting that the EKT approximation to the *N-1* electron wavefunction, Eq. (2), is exact if and only if

$$\langle \Psi_k^{(N-1)} | \left(\hat{H}^{(N-1)} - E_k^{(N-1)}\right)^2 | \Psi_k^{(N-1)} \rangle = 0. \tag{26}$$

Equivalently,

$$\langle \Psi_k^{(N-1)} | \left(\hat{H}^{(N-1)} - \left(E^{(N)} + \varepsilon_k\right)\right)^2 | \Psi_k^{(N-1)} \rangle = 0. \tag{27}$$

Inserting Eq. (2) into Eq. (27) and simplifying the resulting form gives a necessary and sufficient condition for the exactness of the EKT in terms of the two- and three-particle reduced density matrices,

$$0 = \frac{2}{N(N-1)^2} \iiint \begin{bmatrix} \phi_k(\mathbf{z}'_N)\phi_k(\mathbf{z}_N) \\ \times \left(\frac{\nabla^2_{N'}}{2} - v_{ext}(\mathbf{r}'_N) - \varepsilon_k - \frac{N-1}{|\mathbf{r}_{N-1}-\mathbf{r}'_N|}\right)\left(\frac{\nabla^2_N}{2} - v_{ext}(\mathbf{r}_N) - \varepsilon_k - \frac{N-1}{|\mathbf{r}_{N-1}-\mathbf{r}_N|}\right) \\ \times \Gamma_2(\mathbf{z}_{N-1}, \mathbf{z}'_N; \mathbf{z}_{N-1}, \mathbf{z}_N) \end{bmatrix} d\mathbf{z}_{N-1} d\mathbf{z}_N d\mathbf{z}'_N$$

$$+ \frac{3}{N(N-1)^2} \iiint \begin{bmatrix} \phi_k(\mathbf{z}'_N)\phi_k(\mathbf{z}_N) \\ \times \left(\frac{\nabla^2_{N'}}{2} - v_{ext}(\mathbf{r}'_N) - \varepsilon_k - \frac{N-1}{|\mathbf{r}_{N-1}-\mathbf{r}'_N|}\right)\left(\frac{\nabla^2_N}{2} - v_{ext}(\mathbf{r}_N) - \varepsilon_k - \frac{N-1}{|\mathbf{r}_{N-2}-\mathbf{r}_N|}\right) \\ \times \Gamma_3(\mathbf{z}_{N-2}, \mathbf{z}_{N-1}, \mathbf{z}'_N; \mathbf{z}_{N-2}, \mathbf{z}_{N-1}, \mathbf{z}_N) \end{bmatrix} d\mathbf{z}_{N-2} d\mathbf{z}_{N-1} d\mathbf{z}_N d\mathbf{z}'_N$$

(28)

Equation (28) provides a simple and compact test for whether the higher EKT removal energies are exact and, if they are not exact, a measure of their comparative accuracy.

## IV. Summary

The extended Koopmans' theorem (EKT) is a valuable tool for computing ionization energies. If the exact *N*-electron wavefunction is used to formulate the EKT equations (cf. Eqs. (4)-(11)), then the first EKT removal energy, $\varepsilon_0$, and the lowest EKT Dyson orbital, $g_0^{(EKT)}(\mathbf{z})$, are exact. This result can be simply explained by noting that the wavefunction that remains after an electron is moved infinitely far from an *N*-electron system is precisely the lowest-energy wavefunction of the *N-1*-electron ion that can be attained given the spin- and symmetry- constraints inherited from the original *N*-electron system. The EKT works because in the infinite basis-set (alternative, the infinite-grid) limit, the EKT removal orbital, $\phi_0(\mathbf{z})$, is located infinitely far from the molecule in question. Because Eq. (1) is completely general, the EKT is also completely general: the EKT is not limited to molecular systems, fermions, or particles interacting with long-range forces. The same argument used in this paper could be used to establish the exactness of the "generalized" EKT for multiple-ionization.[33]

The EKT is also exact for some other electron removal energies, $\varepsilon_k < \varepsilon_0$. In particular, the preceding asymptotic argument can be used to show that the removal energy to the lowest-energy state of a specified symmetry can sometimes be attained by selecting the appropriate spin- and spatial-symmetry for the EKT removal orbital. Higher-energy states of a given symmetry *might* also be exact, but we do not know of any asymptotic criterion. However, Eq. (28) provides a necessary and sufficient condition for the exactness of any EKT removal energy in terms of the three-particle reduced density matrix.

We conclude with some practical advice for EKT computations. Although one might naïvely expect that diffuse functions are not very important for describing the wavefunction of atomic and molecular cations, it is very important to use diffuse basis sets when computing ionization potentials and Dyson orbitals with EKT. It is also helpful to add a cage of ghost atoms far from the system (but not too far, if the natural orbital occupations corresponding to these basis functions is too small then the EKT equations become numerically unstable), as various linear combinations of the basis functions on these ghost atoms can be used to construct EKT removal orbitals of various symmetries.

**Acknowledgements:** PWA acknowledges support from the Canada Research Chairs, the Alfred P. Sloan foundation, and NSERC. DVF acknowledges support from the research council (BOF) of Ghent University.

**Figure 1**.

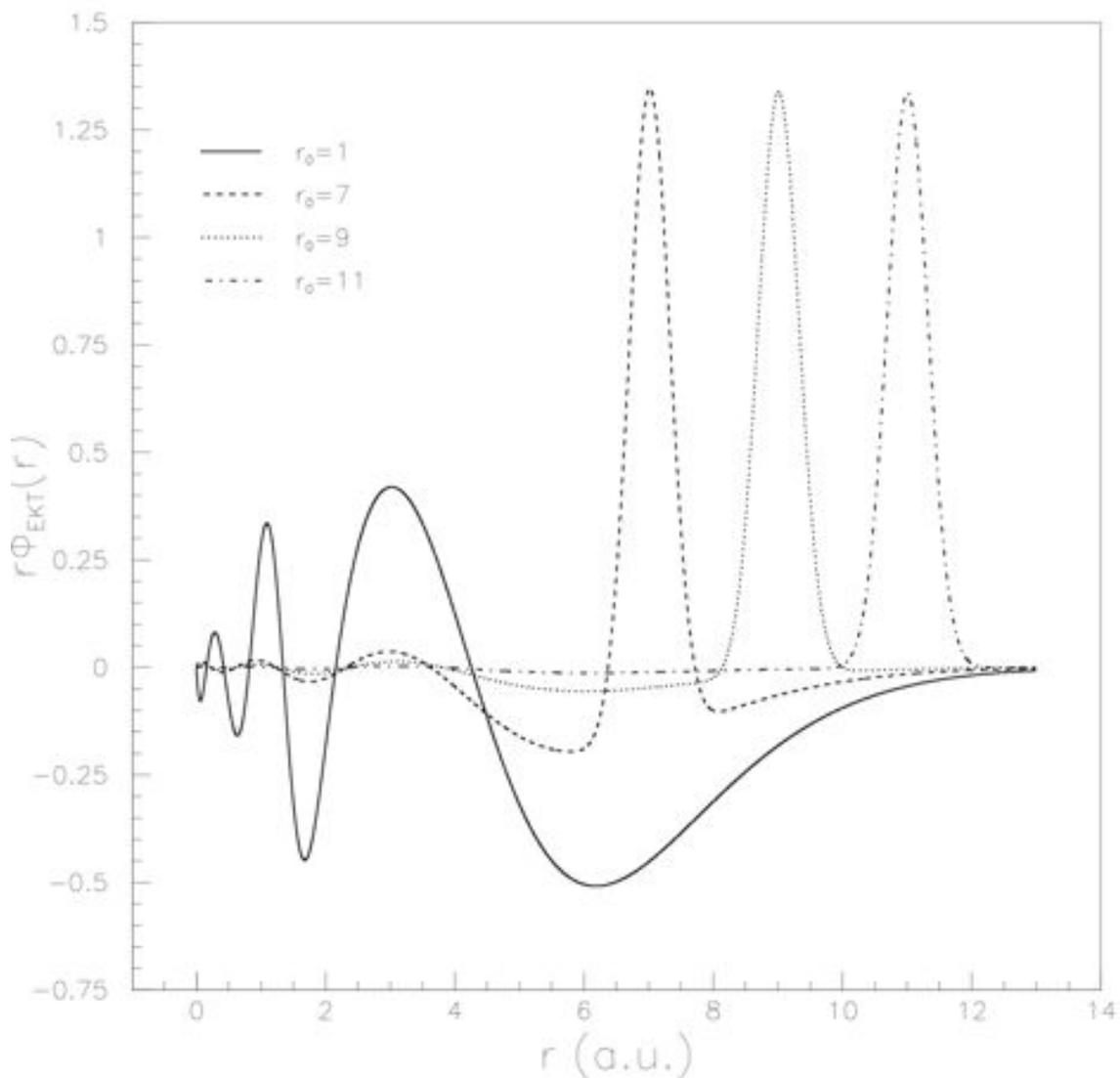

**Caption: The radial wavefunction of the first EKT removal orbital for the Be atom, obtained with a basis consisting of the 6 s-orbitals of the cc-pV5Z basis set and an extra orbital as explained in the text. Atomic units are used. The extra orbital is located at $r_0$ = 1, 7, 9, 11 a.u. Atomic units are used.**

**Figure 2.**

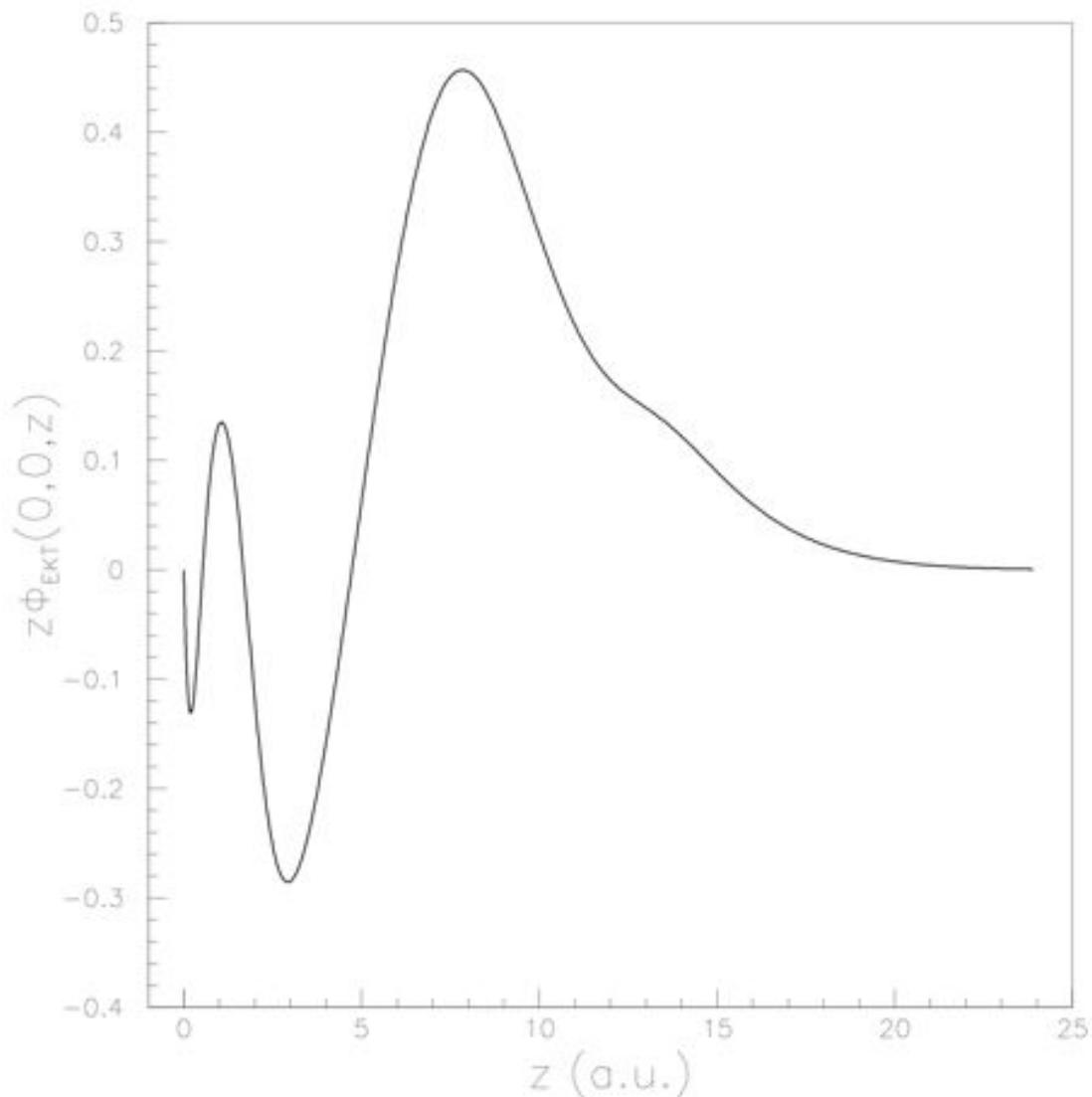

**Caption: The wavefunction of the first EKT removal orbital on the z-axis for the Be atom in the presence of twelve ghost atoms located on the z-axis ±1, ±2, …, ±6 Angstroms from the Be nucleus. The Be atom is described with the 6-311G basis; each ghost atoms is described by a single primitive s-type Gaussian. Atomic units are used.**